\documentclass[aps,pre,showpacs,twocolumn]{revtex4}
\usepackage{bm}
\usepackage{graphicx}
\bibstyle{approve.bib}
\usepackage{amssymb}
\usepackage{amsmath}
\usepackage{esint}
\usepackage{epstopdf}
\usepackage{color}
\usepackage{gensymb}
\usepackage{mathtools}
\usepackage{suffix}

\newcommand{\be}{\begin{equation}}
\newcommand{\ee}{\end{equation}}
\newcommand{\bea}{\begin{eqnarray}}
\newcommand{\eea}{\end{eqnarray}}

\newcommand{\br}{\mathbf{r}}
\newcommand{\bR}{\mathbf{R}}

\newcommand{\bb}{\mathbf{b}}

\newcommand{\bu}{\mathbf{u}}

\newcommand{\bo}{\mathbf{\Omega}}
\newcommand{\bom}{\mathbf{\Omega}}

\newcommand{\e}{\varepsilon}

\newcommand{\pa}{\parallel}

\newcommand{\p}{_{\rm p}}
\newcommand{\Bb}{{\mathbf b}}
\newcommand{\ce}{_{\rm c}}

\newcommand{\s}{_{\rm s}}

\newcommand{\hn}{\hat{\rho}}
\newcommand{\hu}{\hat{u}}

\newcommand{\htt}{\hat{T}}

\newcommand{\bs}{{\mathbf{s}}}

\newcommand{\ew}{\varepsilon_{\rm w}}

\newcommand{\SB}[1]{\textcolor{black} {#1}}
\newcommand{\rf}[1]{\textcolor{black} {#1}}

\begin{document}

\title{Explicit Solvent Effects on Macromolecular Interactions From a Solvent-Augmented Contact Value Theorem}

\author{Sahin Buyukdagli}
\address{Department of Physics, Bilkent University, Ankara 06800, Turkey}

\begin{abstract}

The Derjaguin-Landau-Verywey-Overbeek (DLVO) theory has been a remarkably accurate framework for the characterization of macromolecular stability in water solvent. In view of its solvent-implicit nature neglecting the electrostatics of water molecules with non-negligible charge structure and concentration, the precision of the DLVO formalism is somewhat puzzling. In order to shed light on this issue, we derive from our earlier explicit solvent formalism [S. Buyukdagli el al., Phys. Rev. E, 2013, 87, 063201] a solvent-augmented contact value theorem and assess the contribution of solvent molecules to the interaction of charged membranes. We find that in the case of hydrophobic membranes with fixed charges \SB{embedded in} the membrane surface, the nearly exact cancellation of various explicit solvent effects of substantially large magnitude but opposite sign keeps the  intermembrane pressure significantly close to the double layer force of the DLVO theory. Then, in the case of hydrophilic surface charge groups within the aqueous region, due to the spatial separation of the membrane substrate from the location of the fixed charges where the non-local dielectric response of the structured solvent is sharply localized, the interfacial field energy and the contact charge densities remain unaffected by the explicit solvent. As a result, the hydration of the lipid head groups suppresses the signature of the solvent molecules from the membrane interaction force.

\end{abstract}

\date{\today}

\maketitle

\section{Introduction} 

From nano- to microscale, the equilibrium of macromolecules in aqueous milieu is set by the electrostatic coupling of their omnipresent bound charges. From the stable configuration of like-charged membrane assemblies~\cite{Dub} and densely wrapped DNA molecules around histones~\cite{PodgornikRev} to the cohesion of cement paste~\cite{Jonsson} and colloidal suspensions~\cite{biomatter}, the electrostatic balance of various biological and chemical systems is mainly regulated by the action of short-\SB{range} attractive Van der Waals (vdW)~\cite{Lifshitz,Derja,Tabor1,Tabor2,Ninh,Hough} and \SB{longer-range} repulsive double layer forces~\cite{Levine1,Levine2}. The competition between these opposing forces is precisely at the basis of the DLVO theory.

Although the DLVO formalism has been a significantly efficient tool to characterize the macromolecular stability in water solvent, its theoretical framework includes certain approximations~\cite{DLVO,Isr,ex2}. The first limitation of the DLVO formalism stems from its inclusion of the attractive vdW forces, and the repulsive double layer forces obtained from the Poisson-Boltzmann (PB) theory in an additive fashion. This additivity assumption limited to low macromolecular charges and high salt concentrations has been relaxed by field-theoretic techniques such as one-loop-level weak-coupling (WC) approaches~\cite{PodWKB,Netz1,David1,David2} and electrostatic strong-coupling theories~\cite{Netz2,Netz3,Podgornik2010}.

\rf{In the characterization of the double-layer forces, the relaxation of the fixed surface charge assumption underlying the original DLVO formalism has been another significant step forward. The first incorporation of the surface protonation reactions into the PB theory by Ninham and Parsegian has greatly improved our understanding of the surface force experiments~\cite{pars}. Subsequent works have clarified the generality of the pH-regulated surface charge condition with respect to the constant surface charge constraint~\cite{RudiChReg2}, generalized the charge regulation formalism to hydrogel films~\cite{CruzSoft} and polymer brushes~\cite{LevinChReg}, and extended the theory beyond the mean-field (MF) electrostatic regime~\cite{RudiChReg0,RudiChReg1}. In Ref.~\cite{BuyukpH}, we also integrated the charge regulation mechanism into a unified electrohydrodynamic theory of ion transport and polymer translocation through silicon nitride pores. A consistent review of the literature on charge regulation can be found in Refs.~\cite{Cruz,Rev}.}

The additional limitation of the DLVO approach resides in its solvent-implicit nature common to the classical formulation of electrostatic interactions. In order to relax the assumption of local dielectric response characterizing the solvent-implicit framework, earlier works incorporated structured dielectric permittivity functions into the electrostatic equations of state~\cite{Kor,yar}. Then, within the field-theoretic formulation of electrolyte solutions, explicit solvent has been included in Refs.~\cite{dunyuk,orland1} as point dipoles treated on an equal footing with the salt charges. By generalizing the point-dipole models of Refs.~\cite{dunyuk,orland1} to solvent molecules with finite size, we developed the first field-theoretic formulation of non-local electrostatics able to map from the intramolecular solvent structure to inhomogeneous dielectric response~\cite{PRE1,JCP2013,JCP2014}. \rf{Within the framework of this non-local PB (NLPB) formalism}, we showed that the inclusion of the extended solvent charge structure directly gives rise to the inhomogeneous dielectric permittivity profiles revealed by \rf{atomic force microscopy (AFM)} experiments~\cite{expdiel} and explicit solvent simulations~\cite{Hans,prlnetz}.

The most profound effect brought by \SB{this} non-local dielectric response of structured solvents is a total dielectric void in the vicinity of charge sources. Namely, the aforementioned works revealed that moving away from the membrane at $z=0$ (see Fig.~\ref{fig1}), the relative dielectric permittivity function $\e_{\rm eff}(z)$ assumed to be uniform by the PB theory ($\e_{\rm eff}(z)=\e_{\rm w}\approx78$) actually rises from the vacuum permittivity $\e_{\rm eff}(0)=1$ at the location of the fixed charges to the bulk water permittivity $\e_{\rm eff}(z)\to\e_{\rm w}$ over a few molecular dimensions (see Fig.~\ref{fig2}(a))~\cite{expdiel,Hans,prlnetz,PRE1,JCP2013,JCP2014}.

The corresponding dielectric void at the solvent-membrane interface has a radical consequence on macromolecular interactions. Namely, the absence of this dielectric decrement in the solvent-implicit PB formalism implies that the latter underestimates the interfacial electric field $E(0)\propto\e^{-1}_{\rm eff}(0)$  by a factor of $\e_{\rm w}$ (see Fig.~\ref{fig2}(b)). This means that the attractive component $P_{\rm el}=-\e_{\rm eff}(0)E^2(0)/2$ of the double layer force associated with the interfacial field energy is also underestimated by the DLVO \SB{formalism} by the same factor. Considering that the corresponding error induced by the assumption of local dielectric response is nearly of two orders of magnitude, the experimentally corroborated accuracy of the DLVO \SB {theory} is somewhat puzzling~\cite{Isr,ex2}. 

\SB{In order to shed light on this peculiarity, we carry out the first solvent-explicit characterization of the complex mechanisms responsible for the precision of the DLVO formalism fundamental to our understanding of the macromolecular stability in biological and industrial systems. To this aim, we asses directly the contribution of the explicit solvent molecules to macromolecular interactions between charged membranes. This result is the main achievement of our work.} In Sec.~\ref{mod}, we \SB{first} extend our earlier explicit solvent model~\cite{PRE1} to the case of interfacial membrane charges penetrating the electrolyte region by a finite hydration length. In Sec.~\ref{cvt}, within the framework of this explicit solvent model considered in the MF regime of low membrane charges and monovalent salt, we derive a {\it solvent-augmented contact value theorem} explicitly including the contribution of the solvent molecules to the intermembrane force. In Sec.~\ref{res}, using this generalized contact value theorem, we investigate the effect of the solvent electrostatics on the intermembrane interactions.

In the case of hydrophobic surface charge groups embedded in the membrane substrate, we find that various explicit solvent effects of large magnitude but opposite sign cancel each other out. This drives the net intermembrane pressure significantly close to the double layer force of the DLVO formalism. Then, in hydrophilic membranes carrying charged groups within the aqueous zone, the spatial separation of the non-local dielectric response region at the polar heads from the membrane surface suppresses explicit solvent effects on the individual electrostatic and entropic pressure components. The limitations of our model\SB{, and its potential improvements and applications are elaborated} in Conclusions.

\section{Explicit solvent theory}
\label{mod}

\begin{figure}
\includegraphics[width=1.\linewidth]{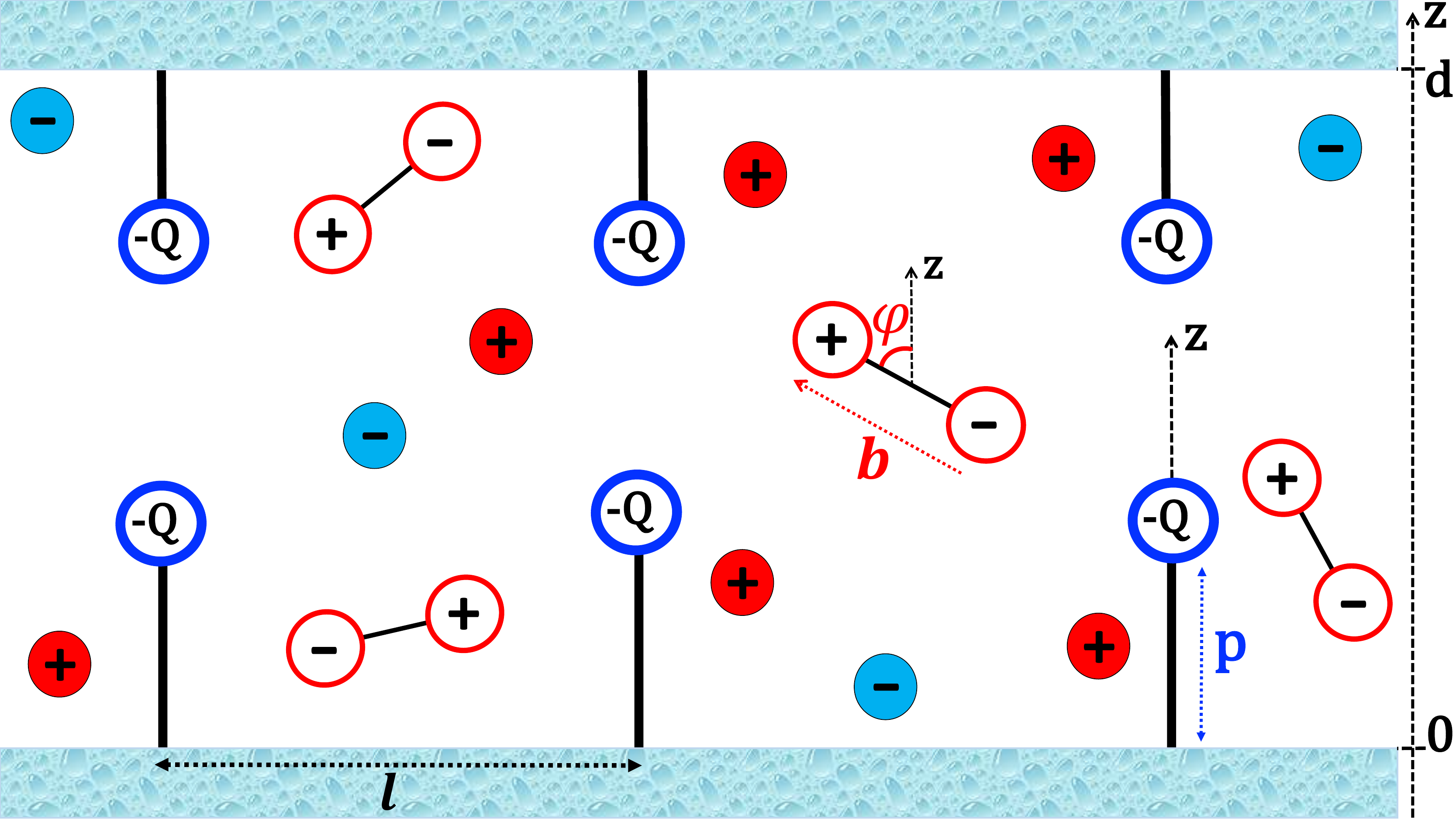}
\caption{(Color online) Schematic depiction of the phospholipid membrane of width $d$ confining $s$ species of monovalent salt ions with bulk concentration $\rho_{i{\rm b}}$,  and dipolar solvent molecules of length $b=1$ {\AA} and reservoir concentration $\rho_{\rm sb}=55$ M. The fixed surface charges $-Q$ of lateral separation distance $l$ are located at the distance $p$ from their wall.}
\label{fig1}
\end{figure}

The composition of the charged system is depicted in Fig.~\ref{fig1}. The liquid is confined between two membrane walls of separation distance $d$. Each membrane wall carries a total of $N$ discretely spaced anionic surface charges $-Q$ of lateral separation distance $l$ and fixed distance $p$ from their wall. The location of the surface charges along the x and y axis of the discrete lattice will be indicated by the indices $n$ and $m$, respectively. 

The electrolyte is composed of dipolar solvent molecules and $s$ species of salt ions \rf{modeled as point charges (see Fig.~\ref{fig1})}. \SB{The rotation of the solvent molecules about their center of mass (C.M.) is characterized by the polar angle $\varphi$ between their dipole vector and the $z$ axis. The confined ions and solvent molecules are in chemical equilibrium with a bulk reservoir located at the ends of the intermembrane region. The ions of the species $i$ have valency $q_i$ and reservoir concentration $\rho_{i{\rm b}}$. Then, each solvent molecule is an overall neutral finite-size dipole composed of a positive and a negative terminal charge. The terminal charges of a given solvent molecule are separated by the fixed distance $b=1$ {\AA}. The solvent concentration in the reservoir is set to the bulk water concentration $\rho_{\rm sb}=55$ M. The bulk dielectric permittivity of the water solvent corresponding to these numerical values follows from the Debye-Langevin equation as $\e_{\rm w}\approx78.3$~\cite{PRE1}.} 

\subsection{Field theoretic partition function}

The dielectric permittivity function of the system reads
\be\label{diel}
\e(\br)=\e_{\rm m}\left[\theta(-z)+\theta(z-d)\right]+\e_0\theta(z)\theta(d-z),
\ee
where $\e_{\rm m}$ and $\e_0=1$ are the relative dielectric permittivity of the membrane and vacuum, respectively. Moreover, the charge density operator is
\bea
\label{ch1}
\hn\ce(\br)&=&\sum_{i=1}^sq_i\sum_{j=1}^{N_i}\delta^3(\br-\bR_{ij})+\sum_{k=1}^{N\s}\htt\left(\br;\bu_k,\bo_k\right)\nonumber\\
&&-Q\sum_{n=1}^N\sum_{m=1}^N\left[\delta^3\left(\br-\br_{nm}\right)+\delta^3\left(\br-\br'_{nm}\right)\right],
\eea
where we defined the mobile ion coordinate $\bR_{ij}$, and the charge structure factor $\htt\left(\br;\bu_k,\bo_k\right)$ of the solvent dipoles with the C.M. coordinate $\bu_k$ and solid angle $\bo_k$. In Eq.~(\ref{ch1}), we also introduced the coordinates of the surface charges defined as
\be
\label{pos1}
\br_{nm}=\bs_{nm}+p\hu_z\;;\hspace{5mm}\br'_{nm}=\bs_{nm}+(d-p)\hu_z,
\ee
with the indices $1\leq n,m\leq N$ and the discrete lateral coordinate $\bs_{nm}=(n-1)l\hu_x+(m-1)l\hu_y$.

The canonical partition function of the system reads
\be
\label{can1}
Z\ce=\prod_{i=1}^s\prod_{j=1}^{N_i}\int\mathrm{d}^3\bR_{ij}\prod_{k=1}^{N\s}\int\frac{\mathrm{d}^2\bo_k}{4\pi}\mathrm{d}^3\bu_k\;e^{-\beta E\ce-\beta E_{\rm n}}.
\ee
In Eq.~(\ref{can1}), we defined the electrostatic interaction energy 
\be\label{can2}
\beta E\ce=\frac{1}{2}\int\mathrm{d^3}\br\mathrm{d^3}\br'\hn\ce(\br)v\ce(\br,\br')\hn\ce(\br')
\ee
including the Coulomb Green's function $v\ce(\br,\br')$ defined in terms of its inverse, $v\ce^{-1}(\br,\br')=-(k_{\rm B}T/e^2)\nabla\cdot\e(\br)\nabla\delta^3(\br-\br')$, where $k_{\rm B}$ is the Boltzmann constant, $T=300$ K the liquid temperature, and $e$ stands for the electron charge. Moreover, we introduced the total steric energy of the charged particles,
\be
\label{can3}
\beta E_{\rm n}=\sum_{i=1}^s\sum_{j=1}^{N_i}V_{\rm n}(\bR_{ij})+\sum_{k=1}^{N\s}V\s(\bu_k,\bo_k),
\ee
with the ionic and solvent onsite potentials $V_{\rm n}(\bR_{\SB{ij}})$ and $V\s(\bu_k,\bo_k)$, respectively.

At this point, we carry out an Hubbard-Stratonovich transformation \SB{to convert the exponential of the electrostatic interaction energy in Eqs.~(\ref{can1})-(\ref{can2}) into a functional integral over} the fluctuating electrostatic potential $\phi(\br)$. Then, in order to account for the chemical equilibrium between the ions and the solvent molecules in the intermembrane region and the reservoir, we pass from the canonical to the grand canonical partition function defined as
\be
\label{can5}
Z_{\rm G}=\prod_{i=1}^s\sum_{N_i=1}^\infty\frac{\lambda_i^{N_i}}{N_i!}\sum_{N\s=1}^\infty\frac{\lambda_s^{N\s}}{N\s!}Z\ce.
\ee
Evaluating the geometric series in Eq.~(\ref{can5}), the partition function takes the functional integral form
\be
\label{can6}
Z_{\rm G}=\int\mathcal{D}\phi\;e^{-\beta H[\phi]}.
\ee
In Eq.~(\ref{can6}), the Hamiltonian functional reads
\bea
\label{can7}
\hspace{-3mm}\beta H[\phi]&=&\frac{k_{\rm B}T}{2e^2}\int\mathrm{d}^3\br\;\e(\br)\left[\nabla\phi(\br)\right]^2-\sum_{i=1}^s\lambda_i\zeta_i-\lambda\s\zeta\s\nonumber\\
&&+iQ\sum_{n=1}^{N}\sum_{m=1}^{N}\left[\phi(\br_{nm})+\phi(\br'_{nm})\right], 
\eea
where the one-body partition functions of the ions and the solvent molecules coupled to the fluctuating background potential $\phi(\br)$ are respectively
\bea
\label{can8}
&&\zeta_i=\int\mathrm{d}^3\br e^{-V_n(\br)+iq_n\phi(\br)},\\
\label{can8II}
&&\zeta\s=\int\frac{\mathrm{d}^2\bo}{4\pi}\mathrm{d}^3\bu\;e^{-V\s(\bu,\bo)+i\int\mathrm{d}^3\br'\htt\left(\br';\bu,\bo\right)\phi(\br')}.
\eea

\subsection{Electrostatic equation of state}

We derive here the MF equation of state \SB{satisfied by the average potential}. Evaluating the saddle-point condition $\delta H[\phi]/\delta\phi(\br)=0$, passing from the complex to the real potential via the transformation $\phi(z)\to i\phi(z)$, and specifying the structure factor of the dipolar solvent molecules characterized by the length vector $\bb$,
\be
\label{can13}
\htt(\br';\bu,\bo)=\delta^3(\br'-\bu-\Bb/2)-\delta^3(\br'-\bu+\Bb/2),
\ee
\SB{the electrostatic equation of state} follows as a three dimensional integro-differential equation,
\bea\label{nlpb0}
&&\frac{k_{\rm B}T}{e^2}\nabla_{\br}\cdot\e(\br)\nabla_{\br}+\sum_{n=1}^s\lambda_nq_ne^{-W_n(\br)-q_n\phi(\br)}\\
&&+\lambda\s\int\frac{\mathrm{d}^2\bom}{4\pi}\left\{e^{-W\s(\br-\bb/2,\bb)}e^{-\phi(\br)+\phi(\br-\bb)}\right.\nonumber\\
&&\hspace{2.2cm}\left.-e^{-W\s(\br+\bb/2,\bb)}e^{-\phi(\br+\bb)+\phi(\br)}\right\}\nonumber\\
&&-Q\sum_{n=1}^{N}\sum_{m=1}^{N}\delta^2(\br_\pa-\bs_{nm})\left[\delta(z-p)+\delta(z-d+p)\right].\nonumber
\eea

Eq.~(\ref{nlpb0}) generalizes the solvent-explicit \SB{NLPB} equation previously derived for continuously distributed interfacial charges on the membrane surface~\cite{PRE1}  to the case of discretely distributed lipid head groups located within the liquid. At this point, by introducing the continuously distributed surface charge approximation, we reduce the dimensionality of the problem. To this aim, we pass from the discrete to the continuous lateral coordinate $\bs_{nm}\to\bs$, and impose in Eq.~(\ref{nlpb0}) the resulting transformation
\be
\sum_{n=1}^{N}\sum_{m=1}^{N}\delta^2(\br_\pa-\bs_{nm})\to\sigma\s\int\mathrm{d}^2\bs\;\delta^2(\br_\pa-\bs)=\sigma\s,
\ee
where we introduced the surface charge density $\sigma\s=l^{-2}$ and the position vector $\br_\pa=x\hu_x+y\hu_y$ along the membrane surfaces. Finally, taking into account the translational symmetry $\phi(\br)=\phi(z)$ associated with the continuous distribution of the interfacial source charges, the extended NLPB Eq.~(\ref{nlpb0}) takes the one-dimensional form 
\bea\label{eq1}
&&\frac{k_{\rm B}T}{e^2}\partial_z\e(z)\partial(z)\phi(z)+\sum_{n=1}^sq_n\SB{\rho_i}(z)+\rho_{{\rm s}+}(z)-\rho_{{\rm s}-}(z)\nonumber\\
&&=\sigma\s Q\left[\delta(z-p)+\delta(z-d+p)\right].
\eea

\rf{In Eq.~(\ref{eq1}), $\rho_i(z)$ stands for the local number density of the salt ions of the species $i$, and $\rho_{{\rm s}+}(z)$ and $\rho_{{\rm s}-}(z)$ correspond to the number densities of the positive and negative terminal charges of the solvent dipoles, respectively. The functional form of these number densities are}
\bea\label{eq2}
\hspace{-3mm}&&\rho_i(z)=\rho_{i{\rm b}}e^{-q_i\phi(z)}\theta\p(z),\\
\label{eq3}
\hspace{-3mm}&&\rho_{{\rm s}+}(z)=\rho_{\rm sb}\theta\p(z)\int_{-b_2(z)}^{b_3(z)}\frac{\mathrm{d}b_z}{2b}e^{-\phi(z)+\phi(z-b_z)},\\
\label{eq3II}
\hspace{-3mm}&&\rho_{{\rm s}-}(z)=\rho_{\rm sb}\theta\p(z)\int_{-b_2(z)}^{b_3(z)}\frac{\mathrm{d}b_z}{2b}e^{-\phi(z-b_z)+\phi(z)},
\eea
with the dipolar projection $b_z=b\cos\varphi$ \SB{onto the $z$ axis}, the steric function $\theta\p(z)=\theta(z)\theta(d-z)$ \rf{including the Heaviside step function $\theta(x)$}, and the integral bounds $b_2(z)={\rm min}\left\{b,d-z\right\}$ and $b_3(z)={\rm min}\left\{b,z\right\}$ taking into account the hard membrane walls impenetrable by the mobile solvent charges. The explicit derivation of the particle densities in Eqs.~(\ref{eq2})-(\ref{eq3II}), and the calculation of the integration boundaries associated with the impenetrable wall condition are explained in Appendix~\ref{a1}. 

From now on, we consider a 1:1 salt solution, and take $s=2$ and $q_\pm=\pm1$. Moreover, we set $Q=1$. The NLPB Eq.~(\ref{eq1}) becomes for $0<z<d$ 
\bea\label{eq1II}
\hspace{-3mm}&&\partial^2_z\phi(z)-\e_{\rm w}\kappa_{\rm i}^2\sinh\left[\phi(z)\right]\\
\hspace{-3mm}&&-\kappa_{\rm s}^2\int_{-b_2(z)}^{b_3(z)}\frac{\mathrm{d}b_z}{2b}\sinh\left[\phi(z)-\phi(z-b_z)\right]\nonumber\\
&&=\frac{2\e_{\rm w}}{\mu}\left[\delta(z-p)+\delta(z-d+p)\right]\nonumber
\eea
\SB{whose numerical solution is explained in Appendix~\ref{rel}}. In Eq.~(\ref{eq1II}), we introduced \rf{the Debye-H\"{u}ckel (DH) screening parameter $\kappa_i^2=8\pi\ell_{\rm w}\rho_{\rm ib}$ quantifying the spatial range of the electrostatic shielding induced by the salt ions, and the solvent screening  parameter $\kappa_s^2=8\pi\ell_{\rm B}\rho_{\rm sb}$ setting the range of the dielectric screening induced by the dipolar solvent molecules~\cite{rem2}. Moreover, the r.h.s. of Eq.~(\ref{eq1II}) includes the Gouy-Chapman (GC) length defined as} $\mu=1/(2\pi \ell_{\rm w}\sigma\s)$, with the Bjerrum length in vacuum \SB{$\ell_{\rm B}=e^2/(4\pi\e_0 k_{\rm B}T)\approx55$ nm} and in water \SB{$\ell_{\rm w}=\ell_{\rm B}/\e_{\rm w}\approx7$ nm}. 

\begin{figure*}
\includegraphics[width=1.\linewidth]{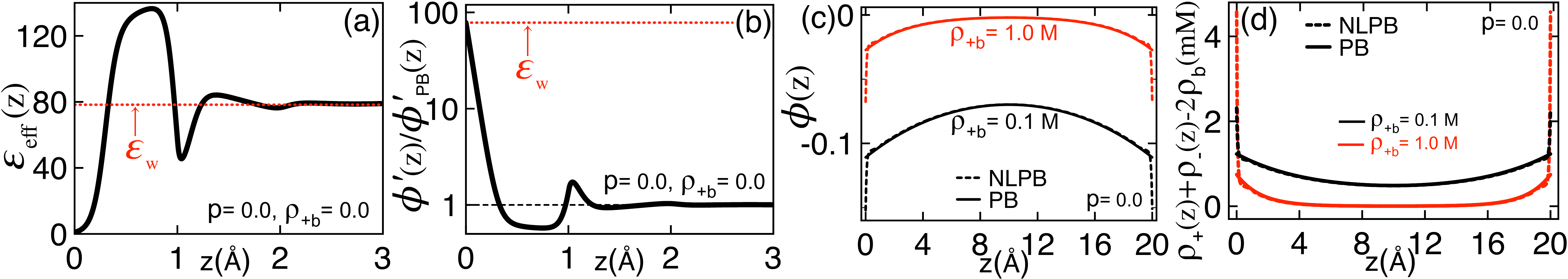}
\caption{(Color online) (a) Effective dielectric permittivity~(\ref{nlep}) and (b) ratio of the NLPB and PB fields in a salt-free liquid. (c) Potential profile and (d) salt excess obtained from the PB theory (solid curves) and the NLPB formalism (dashed curves).}
\label{fig2}
\end{figure*}

\section{Solvent-augmented contact value theorem}
\label{cvt}

In this section, we derive a solvent-explicit contact value theorem relating the intermembrane pressure to the surface value of the mobile charge densities. At the electrostatic MF-level, the grand potential corresponds to the Hamiltonian~(\ref{can7}) evaluated with the solution of the NLPB Eq.~(\ref{eq1}). Accounting for the planar symmetry, the grand potential reads
\bea
\label{fr}
\frac{\beta \Omega_{\rm mf}}{S}&=&-\int_0^d\frac{\mathrm{d}z}{8\pi\ell_{\rm B}}\left[\phi'(z)\right]^2\\
&&-\int_0^d\mathrm{d}z\left\{\rho_+(z)+\rho_-(z)+\rho_{\rm s}(z)\right\}\nonumber\\
&&-\sigma\s\left[\phi(p)+\phi(d-p)\right],\nonumber
\eea
where the C.M. density of the solvent molecules explicitly calculated in Appendix~\ref{a1} is
\be
\label{eq8}
\rho_{\rm s}(z)=\rho_{\rm sb}\int_{-b_1(z)}^{b_1(z)}\frac{\mathrm{d}b_z}{2b}n\s(z,b_z).
\ee
In Eq.~(\ref{eq8}), we used the auxiliary function $b_1(z)={\rm min}\left\{b,2z,2(d-z)\right\}$, and the conditional probability
\be
\label{eq9}
n\s(z,b_z)=e^{-\phi(z+b_z/2)+\phi(z-b_z/2)}.
\ee

The inner pressure follows from the variation of the grand potential~(\ref{fr}) with respect to the membrane separation distance, i.e. $P_{\rm in}=-\delta\Omega_{\rm mf}/\delta(Sd)$. The detailed evaluation of this variation is explained in Appendix~\ref{apr}. Subtracting from the inner pressure its bulk limit \SB{corresponding to the osmotic pressure of the ions and solvent molecules acting on the outer membrane walls, i.e.} $P_{\rm out}=2\rho_{ib}+\rho_{\rm sb}$, the net intermembrane pressure $P_{\rm net}=P_{\rm in}-P_{\rm out}$ follows as
\be\label{eq10}
P_{\rm net}=P_{\rm ion}+P_{\rm sol}+P_{\rm el}.
\ee
In Eq.~(\ref{eq10}), the osmotic pressure components associated with the ionic and solvent entropy excesses are 
\bea
\label{eq10II}
\beta P_{\rm ion}&=&\sum_{i=\pm}\left[\rho_i(d)-\rho_{\rm i b}\right],\\
\label{eq10III}
\beta P_{\rm sol}&=&\rho_{{\rm s}+}(d)+\rho_{{\rm s}-}(d)-\rho_{\rm s b},
\eea
and the electrostatic pressure reads
\be
\label{eq10IV}
\beta P_{\rm el}=\sigma\s \frac{\partial\phi(d-p)}{\partial d}+\frac{\phi'(d^-)^2}{8\pi\ell_{\rm B}}.
\ee
It is noteworthy that the solvent pressure~(\ref{eq10III}) involves the separate contact density of the terminal charges rather than the C.M. density of the solvent molecules.

The solvent-explicit contact value identity in Eq.~(\ref{eq10}) is the main result of this article. In Sec.~\ref{res}, this result will be compared with the PB pressure~\cite{Isr}
\be\label{eq10pb}
P_{\rm net}^{({\rm PB})}=P^{({\rm PB})}_{\rm ion}+P^{({\rm PB})}_{\rm el}
\ee
to characterize explicit solvent effects on macromolecular interactions. We note that in Eq.~(\ref{eq10pb}), the ionic entropy pressure $P^{({\rm PB})}_{\rm ion}$ and the electrostatic pressure $P^{({\rm PB})}_{\rm el}$ follow from Eqs.~(\ref{eq10II}) and~(\ref{eq10IV}) by replacing the potential $\phi(r)$ with the solution of the standard PB equation.

\section{Solvent effects on the interaction of charged membranes}
\label{res}

Here, using the solvent-augmented contact value identity~(\ref{eq10}), we probe the explicit solvent effects on the membrane pressure. To this aim, we investigate separately the case of the surface charge groups located on the membrane surface ($p=0$) and within the liquid ($p>0$). 

The \SB{interaction forces in Eq.~(\ref{eq10})} will be characterized in terms of the dimensionless pressure components
\be
\label{adimp}
{\tilde P}_\alpha=\frac{P_\alpha}{2\pi\ell_{\rm w}\sigma\s^2}
\ee
for $\alpha=\{{\rm net, ion, sol, el}\}$. Moreover, in order to speed up the relaxation algorithm used for the solution of the integro-differential equation~(\ref{eq1II}), we will consider the low membrane charge density $\sigma_{\rm s}=0.01$ ${\rm e}/{\rm nm}^2$ and moderate salt concentrations $\rho_{+{\rm b}}\geq0.1$ M corresponding to the linear DH regime of weak electrostatic potentials.

\subsection{Hydrophobic charge groups ($p=0$)}
\label{hydrophob}

In the case of vanishing tail length $p=0$ corresponding to polar head groups located on the membrane surface, or hydrophobic charge groups partially depleted by water, the electrostatic boundary conditions (BCs) follow from the NLPB Eq.~(\ref{eq1II}) as
\be
\label{bc1} 
\phi'\left(0^+\right)=4\pi\ell_{\rm B}\sigma\s\;;\hspace{5mm}\phi'\left(d^-\right)=-4\pi\ell_{\rm B}\sigma\s.
\ee
Hence, in the same limit $p\to0$ where $\phi'(d^+)=0$ and $\partial_d\phi(d-p)=\phi'(d^-)$, the electrostatic component~(\ref{eq10IV}) of the pressure~(\ref{eq10}) simplifies to
\be
\label{cv2}
\beta P_{\rm el}=-2\pi\ell_{\rm B}\sigma\s^2.
\ee

One notes that the solvent-explicit electrostatic attraction force~(\ref{cv2}) is stronger than the PB prediction~\cite{Isr}
\be\label{prpb}
\beta P^{({\rm PB})}_{\rm el}=-2\pi\ell_{\rm w}\sigma\s^2 
\ee
by the large factor $P_{\rm el}/P_{\rm el}^{({\rm PB})}=\ell_{\rm B}/\ell_{\rm w}=\ew\approx78.3$. The underestimation of the electrostatic interaction force by the PB theory can be explained by accounting for the spatial structure of the effective dielectric permittivity quantifying the inhomogeneous dielectric screening ability of the liquid. The latter can be obtained from the vanishing salt limit of the electric field as~\cite{PRE1,JCP2013}
\be
\label{nlep}
\e_{\rm eff}(z)=\lim_{\kappa_i\to0} \frac{4\pi\ell_{\rm B}\sigma\s}{\phi'(z)}.
\ee
Figs.~\ref{fig2}(a)-(b) display the effective permittivity~(\ref{nlep}) together with the ratio of the NLPB and PB fields. One sees that as a result of the interfacial dielectric void caused by the non-local dielectric response of the explicit solvent liquid, a peculiarity equally observed in AFM experiments~\cite{expdiel} and MD simulations~\cite{Hans,prlnetz} but neglected by the PB theory, the surface field in Eq.~(\ref{bc1}) exceeds the PB field $\phi'_{\rm PB}\left(0^+\right)=4\pi\ell_{\rm w}\sigma\s$ by the factor $\phi'(0)/\phi'_{\rm PB}(0)=\ell_{\rm B}/\ell_{\rm w}=\ew$~\cite{rem1}. This enhances the magnitude of the electrostatic pressure~(\ref{eq10IV}) set by the interfacial field by the same factor. 
\begin{figure*}
\includegraphics[width=1.\linewidth]{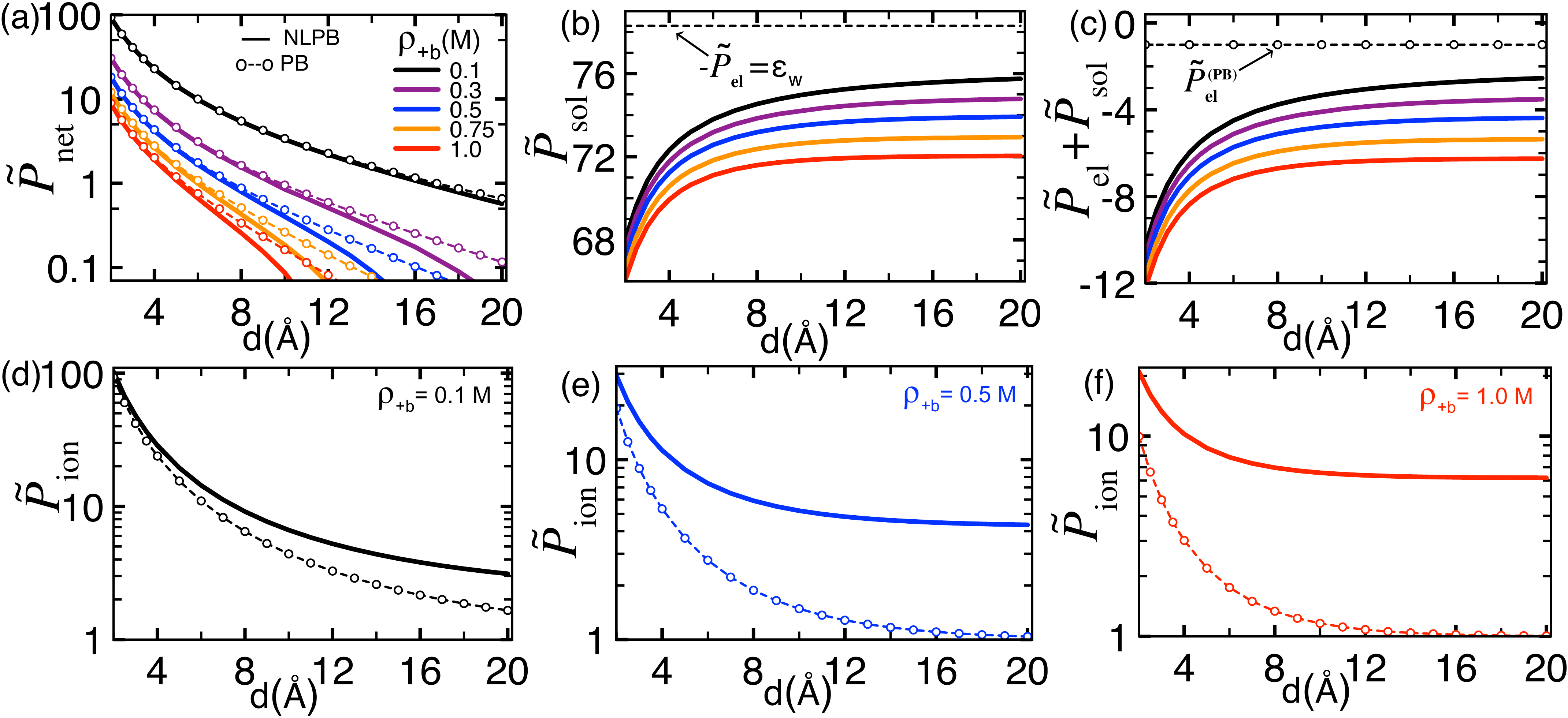}
\caption{(Color online) (a) Intermembrane pressure~(\ref{eq10}) of the NLPB formalism (solid curves) and its solvent-implicit \SB{PB} counterpart~(\ref{eq10pb}) (circles). (b) Repulsive pressure component~(\ref{eq10III}) associated with the entropy of the solvent charges and (c) its sum with the attractive electrostatic pressure~(\ref{cv2}). (d)-(f) Repulsive ionic entropy pressure $P_{\rm ion}$ in Eq.~(\ref{eq10II}) (solid curves) and its implicit solvent counterpart $P^{({\rm PB})}_{\rm ion}$ (circles). The tail length is $p=0$. The salt concentrations are indicated in the legends.}
\label{fig3}
\end{figure*}

In Figs.~\ref{fig3}(a)-(f), we reported the net pressure~(\ref{eq10}) and its components~(\ref{eq10II})-(\ref{eq10IV}) (solid curves), and their implicit-solvent counterparts in Eq.~(\ref{eq10pb}) (circles) versus the intermembrane distance at various salt concentrations indicated in the legends. Fig.~\ref{fig3}(a) shows that in accordance with the agreement of the DLVO theory with the repulsive regime of surface force measurements~\cite{Isr,ex2},  \SB{the PB pressure remains significantly close to the solvent-explicit pressure, i.e. $P^{({\rm PB})}_{\rm net}\approx P_{\rm net}$}. 

Considering that the electrostatic attraction force~(\ref{cv2}) in the explicit solvent is almost two orders of magnitude larger than its solvent-implicit counterpart~(\ref{prpb}), the quasi-overlap of the \SB{PB and NLPB} predictions for the net pressure is puzzling. Below, we show that this seemingly perturbative contribution of the solvent molecules to the intermembrane pressure originates from the cancellation of various explicit solvent effects opposing each other. 

\SB{First, Fig.~\ref{fig3}(b) indicates that the large magnitude of the attractive electrostatic pressure ${\tilde P}_{\rm el}\approx-78.3$ caused by the interfacial dielectric screening deficiency (dashed curve) is largely compensated by the strongly repulsive osmotic pressure component $P_{\rm sol}$ associated with the entropy excess of the solvent molecules (solid curves). However, one also notes that this solvent entropy excess suppressed by salt screening ($\rho_{+{\rm b}}\uparrow P_{\rm sol}\downarrow$) cannot exactly cancel out the amplification of the bare electrostatic pressure by the non-local dielectric response. Consequently, Fig.~\ref{fig3}(c) shows that the explicit solvent-dressed electrostatic pressure remains more attractive than its implicit solvent counterpart,  i.e. $P_{\rm el}+P_{\rm sol}<P_{\rm el}^{({\rm PB})}$. Moreover, unlike the PB-level electrostatic pressure~(\ref{prpb}) unaffected by salt, the magnitude of this attractive force is significantly amplified by salt addition, i.e. $\rho_{+{\rm b}}\uparrow\left[P_{\rm el}+P_{\rm sol}\right]\downarrow$.}

\SB{Interestingly, Fig.~\ref{fig3}(a) shows that the enhancement of the explicit solvent-dressed electrostatic attraction by salt screening is not reflected in the net pressure curves; one notes that at larges salt concentrations, the NLPB pressure is marginally less repulsive than the PB pressure. In order to understand this peculiarity, we now focus on Figs.~\ref{fig2}(c)-(d). Therein, the comparison of the solid and dashed curves shows that the reduced interfacial dielectric permittivity of the explicit solvent amplifies the surface potential and the interfacial salt excess. Figs.~\ref{fig3}(d)-(f) indicate that this leads to the enhancement of the salt-induced repulsive pressure of osmotic origin, i.e. $P_{\rm ion}>P_{\rm ion}^{({\rm PB})}$. As this enhancement compensates almost exactly for the salt-driven amplification of the solvent-dressed attractive force in Fig.~\ref{fig3}(c), i.e.
\be\label{ee1}
P_{\rm ion}-P_{\rm ion}^{({\rm PB})}\approx -\left(P_{\rm el}+P_{\rm sol}-P_{\rm el}^{({\rm PB})}\right), 
\ee
solvent molecules appear to bring a perturbative contribution to the net interaction force displayed in Fig.~\ref{fig3}(a).}

\begin{figure*}
\includegraphics[width=1.\linewidth]{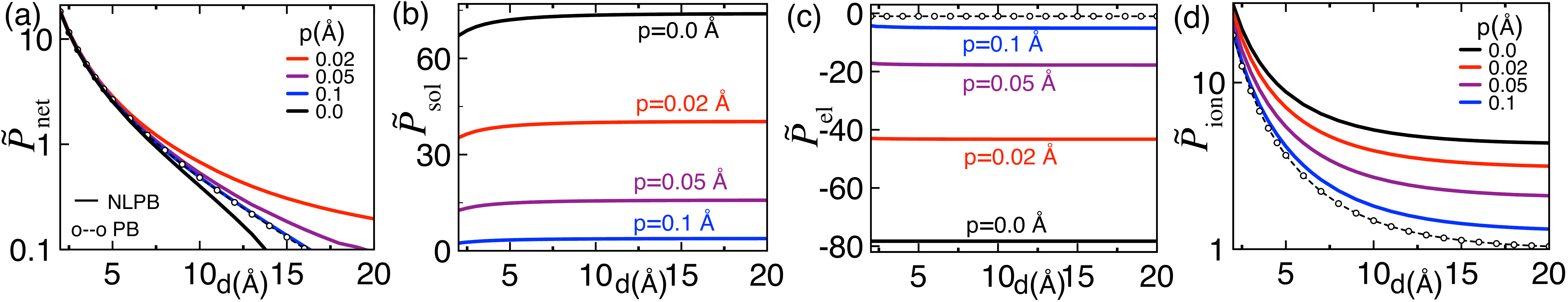}
\caption{(Color online) (a) Net membrane pressure~(\ref{eq10}), (b) solvent-induced pressure~(\ref{eq10III}), (c) electrostatic pressure~(\ref{cv3}), and (d) ionic entropy pressure~(\ref{eq10II}) at various tail lengths $p$ of the surface charge groups. The salt concentration is $\rho_{+{\rm b}}=0.5$ M. Solid curves:  NLPB formalism. Circles: PB formalism.}
\label{fig4}
\end{figure*}

\rf{At this point, we note that our model does not account for the charge structure and the finite size of the ions and the polar heads. However, we emphasize that the ionic Pauling radii of monovalent charges are typically an order of magnitude smaller than the intermediate to large distance regime of Fig.~\ref{fig3} where the pressure components of substantial magnitude and the mechanism driving their mutual compensation are still in effect~\cite{Isr,Frank}. Therefore, although the neglected finiteness of the ion size may quantitatively alter our results, our conclusions are not expected to be qualitatively affected by this approximation. This point is further elaborated in Conclusions.}

\rf{The present MF analysis neglects as well the effect of the electrostatic correlations between the ions and the solvent molecules on the interfacial charge partition. Indeed, the presence of the solid membranes responsible for the non-uniform dielectric and Debye screening induces repulsive image-charge and ionic solvation forces depleting the ions and solvent molecules from the surface~\cite{Netzvdw,Netz1,Netz2,Levin1,Levin2,Buyuk2012,Buyuk2014II}. Thus, these effects expected to reduce the repulsive ionic and solvent pressure components in Fig.~\ref{fig3} may increase the deviation between the PB and NLPB pressures in Fig.~\ref{fig3}(a). However, our earlier theoretical analysis and solvent-implicit MC simulations of inhomogeneous liquids indicate that steric interactions enhance the interfacial particle densities~\cite{jstat,jcp}. Thus, the additional incorporation of the hard-core effects into the present formalism may partially cancel out the aforementioned repulsive forces of electrostatic origin. Therefore, comparisons of our theoretical predictions with the explicit solvent simulations of the present liquid model will be needed to assess accurately the significance of the features neglected in our formalism.}

\begin{figure}
\includegraphics[width=1.\linewidth]{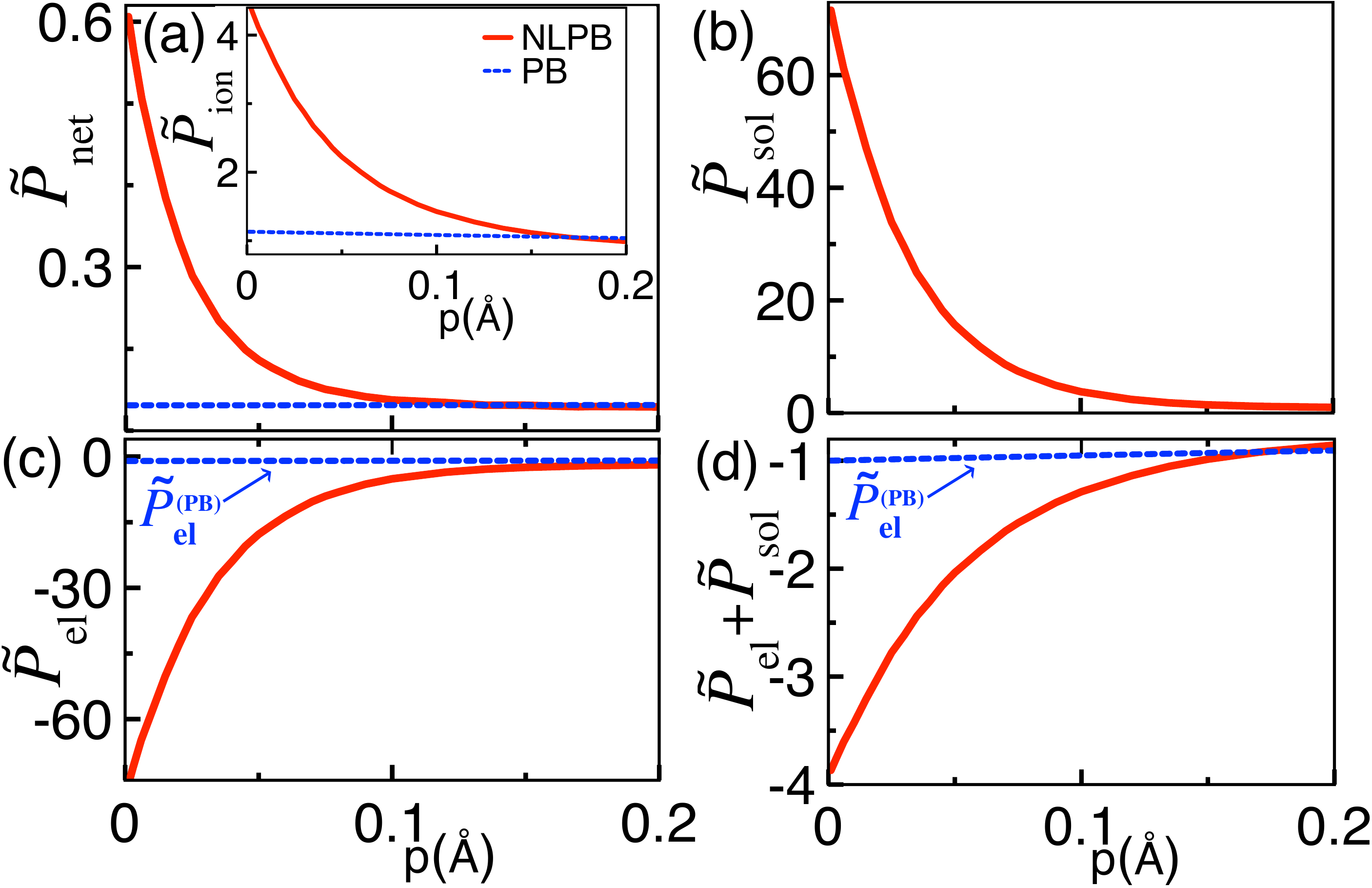}
\caption{(Color online) (a) Net membrane pressure~(\ref{eq10}) (main plot) and ionic entropy pressure~(\ref{eq10II}) (inset), (b) solvent-induced pressure~(\ref{eq10III}), (c) bare electrostatic pressure~(\ref{cv3}), and (d) solvent-dressed electrostatic force versus the tail length $p$ at the membrane separation distance $d=15$ {\AA} and salt concentration $\rho_{+{\rm b}}=0.5$ M. Solid curves:  NLPB formalism. Dashed curves: PB formalism.}
\label{fig5}
\end{figure}

\subsection{Hydrophilic charge groups ($p>0$)}
\label{hydrophil}

In the case of partially hydrophobic polar groups, or equivalently fixed charges embedded on the membrane surface, we showed that the apparent consistency of the PB-level description of intermembrane interactions stems from the mutual cancellation of various solvent-induced effects of opposite sign. Here, we extend this analysis to the case of hydrophilic polar groups, or fixed surface charges penetrating the liquid by an arbitrary length $p$.

\begin{figure*}
\includegraphics[width=1.\linewidth]{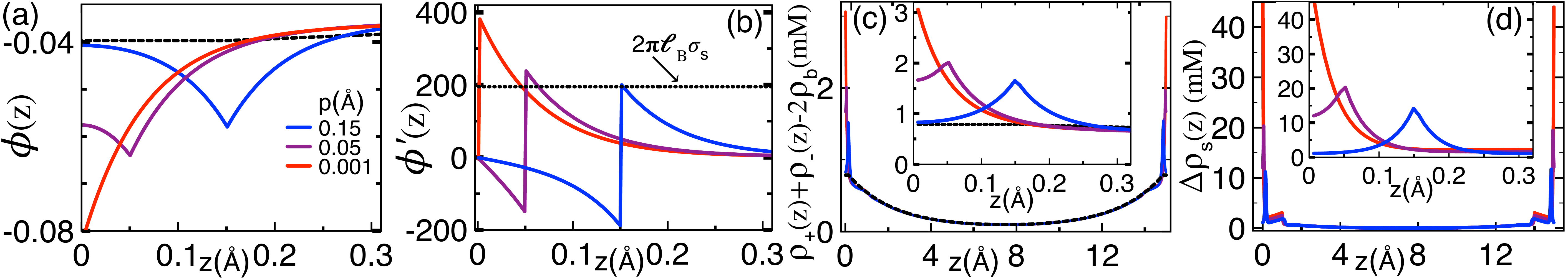}
\caption{(Color online) (a) Interfacial electrostatic potential and (b) field, and excess of (c) salt ion and (d) solvent charge at different tail lengths $p$ indicated in (a). \SB{The solid curves are from the NLPB theory, and the dashed curves in (a) and (c) correspond to the PB formalism. The insets in (c)-(d) display the plots of the main figures at a closer distance from the membrane wall. The salt concentration and the membrane separation distance are $\rho_{+{\rm b}}=0.5$ M and $d=15$ {\AA} in all figures.}} 
\label{fig6}
\end{figure*}
In the specific case of finite tail length ($p>0$) where the surface charge groups are located within the electrolyte, the BCs associated with the NLPB Eq.~(\ref{eq1II}) are
\bea\label{bc2}
&&\hspace{2mm}\phi'(0^+)=\phi'(d^-)=0,\\
\label{bc3}
&&\phi'(z\ce^+)-\phi'(z\ce^-)=4\pi\ell_{\rm B}\sigma\s,
\eea
with $z\ce=\{p,d-p\}$. Using the BCs~(\ref{bc2})-(\ref{bc3}) in Eq.~(\ref{eq10IV}), the electrostatic pressure takes the form
\be
\label{cv3}
\beta P_{\rm el}=\frac{\sigma\s}{2}\left\{\phi'\left[(d-p)^+\right]+\phi'\left[(d-p)^-\right]\right\}.
\ee

Fig.~\ref{fig4} displays the net pressure profile and its components at various tail lengths $p$. One sees that as the tail length rises, the repulsive force $P_{\rm sol}$ associated with the solvent entropy decays rapidly, while the magnitude of the attractive electrostatic pressure $P_{\rm el}$ and the repulsive ionic entropy pressure $P_{\rm ion}$  (solid curves) amplified by the non-local dielectric response drop to their PB limit (circles).  Due to this hydration-driven suppression of the \SB{explicit} solvent effects on the individual force components, the net intermembrane force displayed in Fig.~\ref{fig4}(a) quickly converges to the PB pressure.

In Fig.~\ref{fig5}, we illustrate the alteration of the individual force components by the tail length $p$ at \SB{the fixed intermembrane distance $d=15$ {\AA}}. In accordance with Fig.~\ref{fig4},  the plots indicate that the increase of the tail length causes the uniform \SB{and} rapid decay of the solvent-driven \SB{osmotic} force $P_{\rm sol}$, and the convergence of the remaining force components to their PB limit. As a result, Fig.~\ref{fig5}(a) shows that the moderate repercussion of the non-local dielectric response on the net interaction force disappears rapidly over the short penetration length $p\approx0.1$ {\AA}.

With the aim to shed light on the mechanism behind the removal of the explicit solvent effects on the individual force components by the full hydration of the surface charge groups, in Fig.~\ref{fig6}, we reported the profile of the average potential, the interfacial field, the salt excess, and the solvent excess density
\be
\Delta\rho_{\rm s}(z)=\rho_{{\rm s}+}(z)+\rho_{{\rm s}-}(z)-\rho_{\rm sb}\frac{b_2(z)+b_3(z)}{2b}.
\ee

We first focus on the repulsive \SB{osmotic pressure} components. Figs.~\ref{fig6}(a) and (c)-(d) show that the penetration of the surface charge groups into the liquid moves the explicit solvent-enhanced potential region and the corresponding peak of the particle densities (solid curves) away from the solid membrane surface. \SB{The corresponding splitting of the membrane wall and the zone governed by the non-local dielectric response drives the surface potential and the contact charge densities fixed by the latter towards their PB limit (dashed curves). As a result, the ionic and solvent pressures in Figs.~\ref{fig4}(b) and (d) set by these contact charge densities are reduced by the full hydration of the polar groups down to their PB value.}

In order to understand the decay of the attractive electrostatic pressure in Fig.~\ref{fig4}(c) by hydration, we note that the use of the potential symmetry $\phi(d-z)=\phi(z)$ together with the BC~(\ref{bc3}) allows to recast Eq.~(\ref{cv3}) as
\be
\label{cv4}
\beta P_{\rm el}=-\sigma\s\left[\phi'(p^+)-2\pi\ell_{\rm B}\sigma\s\right].
\ee 
The interfacial field profile in Fig.~\ref{fig6}(b) shows that as the surface charge groups penetrate the liquid, due to the interaction of the polar heads with the salt solution on their two sides, the electric field peak $\phi'(p^+)$ reduced by almost a factor of two drops towards the value of $2\pi\ell_{\rm B}\sigma\s$. However, as $p\to\infty$, the field $\phi'(p^+)$ remains weakly above this limit value such that their difference in Eq.~(\ref{cv4}) tends to the PB value of the electrostatic pressure. This explains the rapid convergence of the electrostatic force in Fig.~\ref{fig4}(c) to its PB limit.

\rf{At this point, we note that the relaxation of the fixed surface charge condition at the basis of our theory would provide a significant extension of the present model. As discussed in the Introduction, the surface protonation effects governing the macromolecular charge dynamics from low to moderate pH conditions have been intensively studied by MF approaches~\cite{pars,CruzSoft,RudiChReg2,BuyukpH} and correlation-corrected theories~\cite{RudiChReg0,RudiChReg1,LevinChReg}. We emphasize that due to the solvent-explicit nature of our formalism, this extension will require the coupling of the solvent component of our electrolyte model to the protonation reactions in an explicit fashion.}

\rf{The steric and hydration interactions between the surface charges of the adjacent membrane walls are additional features relevant to the case of hydrophilic polar groups studied herein. Previous explicit solvent simulations showed that these effects can bring a repulsive contribution to the intermembrane interactions~\cite{hyd1,hyd2,hyd3}. It should be however noted that the incorporation of these features into the present formalism will necessitate the inclusion of surface specific effects regulating the level of the hydration experienced by the charge groups.}

\section{Conclusions}

The consistent characterization of macromolecular interactions in water solvent is an essential step towards understanding the stability of biological systems. In this article, we derived \SB{the first} contact value theorem explicitly including the contribution of \SB{structured} solvent molecules to the intermembrane force. Within this solvent-explicit framework, we analyzed at the MF-level the effect of solvent electrostatics on the interaction of charged membranes carrying hydrophobic and hydrophilic surface charge groups.

In the case of hydrophobic membranes characterized by polar head groups embedded in the membrane surface, the interfacial dielectric screening deficiency originating from the non-local dielectric response of the explicit solvent amplifies the surface field and the \SB{attractive electrostatic force set by the latter} by a factor of $\e_{\rm w}\approx78.3$. However, \SB{the resulting amplification of the interfacial potential associated with this surface field} also increases the interfacial salt and solvent charge excesses, and enhances the \SB{corresponding osmotic repulsive forces}. Due to the almost exact cancellation of these effects with substantial magnitude but opposite sign, \SB{the role of the explicit solvent in membrane interactions appears to be negligible.}

For hydrophilic membranes, the penetration of the polar head groups into the liquid \SB{shifts the explicit solvent-enhanced interfacial field and potential, and the resulting peak of the salt and solvent densities away from the membrane surface. This reduces} the attractive pressure component, and the repulsive \SB{osmotic pressures} set by the contact particle densities separately to their implicit solvent limit. As a result, \SB{the full hydration of the fixed surface charge groups} drives the net intermembrane pressure to the double-layer force of the \SB{DLVO} formalism.

In the present study, \SB{we carried-out the first solvent-explicit analysis of the mechanisms behind the experimentally corroborated accuracy of the solvent-implicit DLVO theory. This point is the main achievement of our work. As any new approach, our formalism has limitations that can be gradually relaxed by future works.} 

\SB{First,} we investigated the predictions of the solvent-augmented contact value theorem within the DH regime of weakly charged membranes and strong salt. Future works can extend our analysis to the non-linear GC regime of dilute salt~\cite{Buyuk2015} or even relax the present MF approximation by including charge correlations~\cite{JCP2014}. 

\rf{An additional system feature neglected in our formalism is the finite ion size. As our explicit solvent theory includes the salt charges and the solvent molecules on an equal footing, the characteristic ion size relevant to our model is the Pauling radius corresponding to the bare radius of the salt charges without their hydration shell. The typical Pauling radii for monovalent ions are on the order of $1$ {\AA}~\cite{Isr,Frank}. It should be noted that the corresponding ion sizes are an order of magnitude smaller than the intermediate to large distance branch of the pressure curves in Fig.~\ref{fig3} where the large magnitude of the pressure components and the complex mechanism responsible for their mutual cancellation remain intact.  Indeed, the finite ion size setting the closest approach distance of the ions to the membrane surface is expected to increase the distance scale of the pressure curves in Fig.~\ref{fig3} by a few Pauling radii. Therefore, the introduction of the finite Pauling radii alone is not expected to alter qualitatively the predictions of Sec.~\ref{hydrophob}. Although comparisons with numerical simulations of dipolar liquids are certainly required to confirm these points, we emphasize that the present formalism, and its point-dipole and implicit solvent versions neglecting the finite ion size have been able to reproduce with reasonable accuracy experimental trends driven by salt charges, such as the salt-induced dielectric decrement in bulk electrolytes~\cite{Buyuk2022}, the salt dependence of the differential capacitance of low-permittivity materials~\cite{Buyuk2012}, and the ion conductivity of strongly confined $\alpha$-Hemolysin channels and solid-state nanopores~\cite{Buyuk2014}.  It should be also added that the incorporation of the spherical charge distribution of the salt ions would break the planar symmetry of the system, requiring the analysis of the model via the solution of the full three-dimensional integro-differential Eq.~(\ref{nlpb0}). This formidable task of tremendous technical complexity is beyond the scope of the present work.}

\rf{We also emphasize that in Figs.~\ref{fig4}-\ref{fig6}, the significantly small values of the characteristic penetration lengths $p$ are expected to be quantitatively affected by the consideration of additional system features neglected in our model, such as the finite size of the polar heads, and the ion size effect elaborated above. It should be however noted that our comprehensive study provides the first insightful conclusion on the removal of the explicit solvent effects upon the full hydration of the surface charge groups. Thus, this relevant result is the main achievement of Sec.~\ref{hydrophil}, and the analysis presented therein should not be considered as an attempt to identify with quantitative precision the critical hydration lengths where explicit solvent effects dissipate.}

\SB{Future works should incorporate into the present model} additional relevant features of \SB{confined} charged liquids, such as the \SB{formation of hydrogen bonds} between the water molecules, the incompressibility of the water solvent, \rf{the surface protonation driving the dynamics of the macromolecular charges in the low to moderate pH regime~\cite{pars,RudiChReg2,CruzSoft,LevinChReg,RudiChReg0,RudiChReg1,BuyukpH,Cruz,Rev}, electrostatic correlations inducing repulsive solvation and image-charge forces~\cite{Netzvdw,Levin1}, and hydrophobic image-dipole interactions~\cite{Buyuk2014II}. In particular, via the introduction of the ionic polarizability within the Drude oscillator model, we are currently working on the relaxation of the point charge approximation. This work will be presented in a separate article.} \SB{A through} confrontation of our theoretical predictions with solvent-explicit MD simulations will be equally needed to asses the significance of the aforementioned effects on the electrostatic stability of macromolecules.

\SB{Before concluding, it is noteworthy that unlike the surface salt excess of the PB theory reduced by salt addition to the reservoir (solid curves in Fig.~\ref{fig2}(d)), the interfacial salt excess in the explicit solvent liquid is strongly amplified by the increment of the bulk salt concentration (dashed curves). This peculiarity responsible for the salt-induced enhancement of the deviation between the ionic PB and NLPB pressures in Figs.~\ref{fig3}(d)-(f) may have a substantial effect on the ion conductivity of nanofluidic devices. Therefore, we would like to probe the role played by explicit solvent on nano-confined charge transport in a future work.} 

\smallskip
\appendix

\section{Derivation of the charge densities}
\label{a1}

We derive here the local densities of the charged particles, and the orientational probabilities of the surface and solvent dipoles.

\subsection{Salt ions}

The MF-level salt ion density follows from the thermodynamic relation $\rho_i(\br)=-\delta\ln Z_{\rm G}/\delta V_i(\br)\approx\delta\left(\beta H\right)/\delta V_i(\br)$ as $\rho_i(\br)=\lambda_ie^{-V_i(\br)+iq_i\phi(\br)}$. Noting that the bulk region $\br\to\infty$ is characterized by vanishing potentials $V_i(\br)=\phi(\br)=0$, the ion fugacity follows as $\lambda_i=\rho_{i{\rm b}}$. Passing to the real potential via the transformation $\phi(\br)\to i\phi(\br)$, accounting for the planar symmetry $\phi(\br)=\phi(z)$, and imposing the steric constraint associated with the ion confinement in the intermembrane region, the density of the salt ions becomes
\be
\label{a1}
\rho_i(z)=\rho_{i{\rm b}}\theta\p(z)e^{-q_i\phi(z)},
\ee
with the auxiliary function
\be\label{a2}
\theta\p(z)=\theta(z)\theta(d-z).
\ee

\subsection{Solvent particles}

In order to derive the solvent number density, we define the steric solvent potential in Eq.~(\ref{can8II}) as $V\s(\bu,\bom)=V_{{\rm s,}+}(\br+\Bb/2)+V_{\rm s,cm}(\br)+V_{{\rm s,}-}(\br-\Bb/2)$, where the potentials $V_{{\rm s}}(\br)$ with $i=\left\{{\pm,{\rm cm}}\right\}$ act on the terminal charges and the C.M. of the molecule. Using the thermodynamic identity $\rho_{{\rm s}i}(\br)=\delta\left(\beta H\right)/\delta V_{{\rm s,}i}(\br)$, imposing the plane symmetry together with the impenetrability of the membrane walls, and introducing the dipolar projection variable $b_z=b\cos\varphi$, the number and charge densities of the solvent molecules follow as
\bea
\label{a3}
\hspace{-3mm}\rho_{\rm s}(z)&=&\rho_{\rm sb}\int_{-b_1(z)}^{b_1(z)}\frac{\mathrm{d}b_z}{2b}n\s\left(z,b_z\right),\\
\label{a4}
\hspace{-3mm}\rho_{\rm s+}(z)&=&\rho_{\rm sb}\theta\p(z)\int_{-b_2(z)}^{b_3(z)}\frac{\mathrm{d}b_z}{2b}n\s\left(z-b_z/2,b_z\right),\\
\label{a5}
\hspace{-3mm}\rho_{\rm s-}(z)&=&\rho_{\rm sb}\theta\p(z)\int_{-b_3(z)}^{b_2(z)}\frac{\mathrm{d}b_z}{2b}n\s\left(z+b_z/2,b_z\right),
\eea
with the conditional probability
\be
\label{a6}
n\s(z,b_z)=e^{-c_+\phi(z+b_z/2)+c_-\phi(z-b_z/2)},
\ee
and the auxiliary functions
\bea\label{a7}
b_1(z)&=&{\rm min}\left\{b,2z,2(d-z)\right\},\\
\label{a8}
b_2(z)&=&{\rm min}\left\{b,d-z\right\},\\
\label{a9}
b_3(z)&=&{\rm min}\left\{b,z\right\}.
\eea

\section{Relaxation algorithm for the solution of the NLPB Eq.~(\ref{eq1II})}
\label{rel}

We introduce here a relaxation algorithm for the numerical solution of the NLPB Eq.~(\ref{eq1II}) on a one-dimensional discrete grid. The $2N+1$ nodes of the grid will be labeled by the index $n$ defined in the interval $1\leq n\leq 2N+1$. To this aim, we pass from the continuous variable $z$ to the discrete lattice coordinate $z_n=\delta(n-1)$, with the grid spacing defined as $\delta=d/(2N)$. Using the finite difference definition of derivatives, Eq.~(\ref{eq1II}) can be cast in the following discrete form 
\bea
\label{d1}
\phi_n&=&\frac{1}{2}\left\{\phi_{n+1}+\phi_{n-1}-r_i\sinh\phi_n\right.\\
&&\left.\hspace{5mm}-r\s\sum_{j=-j_2(n)}^{j_3(n)}\sinh\left[\phi_n-\phi_{n-j+1}\right]\right\}.\nonumber
\eea
In Eq.~(\ref{d1}), the potential values on the lattice nodes are defined as $\phi_n\equiv\phi(z_n)$. Moreover, we introduced the coefficients $r\s=\delta^3\kappa\s^2/(2b)$ and $r_i=\delta^2\ew\kappa_i^2$, the functions $j_2={\rm min}(n_b,2N-n+2)$ and $j_3(n)={\rm min}(n_b,n)$, and the index $n_b={\rm int}(b/\delta+1)$. 

Eq.~(\ref{d1}) should be solved recursively by injecting at the first iterative step a guess potential profile into the r.h.s., and using the output solution as the updated input potential at the next iterative step. This cycle should be continued until numerical convergence is achieved. The key requirement for convergence is the injection of an adequate guess potential at the first iterative step. We found that the standard solvent-implicit PB solution can be used as an input potential exclusively if i) Eq.~(\ref{d1}) is evaluated by starting at the mid-pore at $z=d/2$ and moving to the membrane surface at $z=0$, and ii) the input potential is updated not only at the end of each iterative cycle but also at each node $n$ of the grid during the cycle.

During the execution of the relaxation scheme, the potential values at $z>d/2$ can be obtained by exploring the mirror symmetry $\phi(z)=\phi(d-z)$ implying 
\be\label{d2}
\phi_{n>N+1}=\phi_{2N-n+2}.
\ee
Moreover, on the membrane surface at $n=1$, the r.h.s. of Eq.~(\ref{d1}) requires the value of $\phi_0$. This can be determined by using the discrete form of the BCs~(\ref{bc1}) and~(\ref{bc2}). For $p=0$, the discretization of the BC~(\ref{bc1}) gives
\be\label{d3}
\phi_0=\phi_2-\frac{4\delta\ew}{\mu}.
\ee
Then, for $p>0$, the BC~(\ref{bc2}) yields $\phi_0=\phi_2$. Finally, we note that the discretization of the BC~(\ref{bc3}) yields
\be\label{d4}
\phi_{n_p}=\frac{1}{2}\left(\phi_{n_p+1}+\phi_{n_p-1}\right)-\frac{\delta\ew}{\mu}.
\ee
Thus, at $n=n_p$, one should use for $p>0$ Eq.~(\ref{d4}) instead of Eq.~(\ref{d1}).

\section{Derivation of the solvent-explicit contact value theorem~(\ref{eq10})}
\label{apr}

In this appendix, we present the derivation of the solvent-explicit contact value theorem~(\ref{eq10}). To this aim, we first note that at the MF level, the thermodynamic potential of the system corresponds to the Hamiltonian~(\ref{can7}) evaluated with the solution of the saddle-point Eq.~(\ref{eq1}). Taking into account the planar symmetry, the MF grand potential reads
\bea
\label{a16}
\frac{\beta \Omega_{\rm mf}}{S}&=&-\int_0^d\frac{\mathrm{d}z}{8\pi\ell_{\rm B}}\left[\phi'(z)\right]^2\\
&&-\int_0^d\mathrm{d}z\left\{\rho_+(z)+\rho_-(z)+\rho_{\rm s}(z)\right\}\nonumber\\
&&-\sigma\s Q_-\left[\phi(p)+\phi(d-p)\right].\nonumber
\eea

The pressure acting on the inner membrane walls corresponds to the variation of the thermodynamic potential~(\ref{a16}) with respect to the intermembrane distance, i.e. $P_{\rm in}=-\delta\Omega_{\rm mf}/\delta(Sd)$. Using the Leibniz rule, this yields
\bea\label{a17}
P_{\rm in}&=&-\frac{1}{S}\int_{-\infty}^{+\infty}\mathrm{d}z\frac{\delta\Omega_{\rm mf}}{\delta\phi(z)}\frac{\partial\phi(z)}{\partial d}+\frac{\phi'(d^-)^2}{8\pi\ell_{\rm B}}+\sum_{i=1}^s\rho_i(d)\nonumber\\
&&+\sigma\s Q_-\frac{\partial\phi(d-p)}{\partial d}+\rho_{\rm s}(d)+\int_0^d\mathrm{d}z\;\partial_d\rho_{\rm s}(z).
\eea

We first note that as the average potential $\phi(z)$ satisfies the saddle-point condition $\delta\Omega_{\rm mf}/\delta\phi(z)=0$, the first term on the r.h.s. of Eq.~(\ref{a17}) vanishes. Then, Eqs.~(\ref{a3}) and~(\ref{a7}) show that as a result of the steric constraint at the impenetrable wall, the contact density of the solvent molecules corresponding to the fifth term on the r.h.s. of  Eq.~(\ref{a17}) vanishes as well, i.e. $b_1(d)=0$ and $\rho_{\rm s}(d)=0$. Consequently, the inner pressure~(\ref{a17}) simplifies to
\bea\label{a18}
P_{\rm in}&=&\sum_{i=1}^s\rho_i(d)+\sigma\s Q_-\frac{\partial\phi(d-p)}{\partial d}+\frac{\phi'(d^-)^2}{8\pi\ell_{\rm B}}\nonumber\\
&&+\int_0^d\mathrm{d}z\;\partial_d\rho_{\rm s}(z).
\eea

In order to evaluate the integral term in Eq.~(\ref{a18}), we note that according to Eqs.~(\ref{a3}) and~(\ref{a7}), $\partial_d\rho_{\rm s}(z)\neq0$ if $b_1(z)=2(d-z)$, which holds if $2(d-z)<2z$ and $2(d-z)<b$, or $z>d/2$ and $z>d-b/2$. Since $b<d$, the derivative $\partial_d\rho_{\rm s}(z)$ is therefore finite only in the interval $d>z>d-b/2$. This remark allows us to express the integral term of Eq.~(\ref{a18}) as
\bea
\label{a19}
&&\int_0^d\mathrm{d}z\;\partial_d\rho_{\rm s}(z)\\
&&=\rho_{\rm sb}\int_0^d\mathrm{d}z\;\theta(z-d+b/2)\partial_d\int_{-2(d-z)}^{2(d-z)}\frac{\mathrm{d}b_z}{2b}n\s(z,b_z)\nonumber\\
\label{a20}
&&=\frac{\rho_{\rm sb}}{b}\int_{d-b/2}^d\mathrm{d}z\left\{e^{-\phi(d)+\phi(2z-d)}+e^{-\phi(2z-d)+\phi(d)}\right\},\nonumber\\
\eea
where the second equality followed from the use of \SB{Eqs.~(\ref{a3}) and~(\ref{a6})}, and the Leibniz integral rule. Finally, introducing in Eq.~(\ref{a20}) the change of variable $z\to d-b_z/2$, after some algebra, one can recast the integral as the sum of the solvent charge densities~(\ref{a4})-(\ref{a5}), i.e.
\be
\label{a21}
\int_0^d\mathrm{d}z\;\partial_d\rho_{\rm s}(z)=\rho_{{\rm s}+}(d)+\rho_{{\rm s}-}(d).
\ee
Substituting Eq.~(\ref{a21}) into Eq.~(\ref{a18}), and subtracting from the latter the bulk pressure $P_{\rm out}=\sum_i\rho_{ib}+\rho_{\rm sb}$, the net intermembrane pressure $P=P_{\rm in}-P_{\rm out}$ follows in the form of the solvent-explicit contact value theorem \SB{corresponding to Eq.~(\ref{eq10}) in the main text, i.e.}
\bea
\label{a22}
\beta P&=&\sigma\s Q_-\frac{\partial\phi(d-p)}{\partial d}+\frac{\phi'(d^-)^2}{8\pi\ell_{\rm B}}\\
&&+\sum_{i=1}^s\left[\rho_i(d)-\rho_{\rm i b}\right]+\rho_{{\rm s}+}(d)+\rho_{{\rm s}-}(d)-\rho_{\rm s b}.\nonumber
\eea

\end{document}